\documentclass[superscriptaddress,aps,prb,longbibliography,twocolumn]{revtex4-1}

\usepackage{graphicx}
\usepackage[colorlinks=True,linkcolor=red,citecolor=blue,urlcolor=blue]{hyperref}
\usepackage[version=4]{mhchem}
\usepackage{siunitx}
\usepackage{multirow}
\usepackage{bm}
\usepackage{amssymb}
\usepackage{array}
\usepackage{makecell}
\usepackage{braket}
\usepackage{booktabs}
\usepackage{dcolumn}

\newcounter{Methods}

\newcounter{SI}

\usepackage{etoolbox}
\makeatletter         
\newcounter{firstbib} 

\AtBeginEnvironment{thebibliography}{
	
}

\apptocmd{\thebibliography}{
	\setcounter{NAT@ctr}{\value{firstbib}} 
	
}{}{}  

\AtEndEnvironment{thebibliography}{
	\setcounter{firstbib}{\value{NAT@ctr}}
	
}
\makeatother

\begin{document}

\title{Effect of uniaxial stress on the electronic band structure of NbP}

\author{Clemens Schindler}
\email{clemens.schindler@cpfs.mpg.de}
\affiliation{Max Planck Institute for Chemical Physics of Solids, 01187 Dresden, Germany}
\affiliation{Institut f\"{u}r Festk\"orper- und Materialphysik, Technische Universit\"at Dresden, 01062 Dresden, Germany}

\author{Jonathan Noky}
\affiliation{Max Planck Institute for Chemical Physics of Solids, 01187 Dresden, Germany}

\author{Marcus Schmidt}
\affiliation{Max Planck Institute for Chemical Physics of Solids, 01187 Dresden, Germany}

\author{Claudia Felser}
\affiliation{Max Planck Institute for Chemical Physics of Solids, 01187 Dresden, Germany}

\author{Jochen Wosnitza}
\affiliation{Institut f\"{u}r Festk\"orper- und Materialphysik, Technische Universit\"at Dresden, 01062 Dresden, Germany}
\affiliation{Hochfeld-Magnetlabor Dresden (HLD-EMFL) and W\"urzburg-Dresden Cluster of Excellence ct.qmat, Helmholtz-Zentrum Dresden-Rossendorf, 01328 Dresden, Germany}

\author{Johannes Gooth}
\email{johannes.gooth@cpfs.mpg.de}
\affiliation{Max Planck Institute for Chemical Physics of Solids, 01187 Dresden, Germany}
\affiliation{Institut f\"{u}r Festk\"orper- und Materialphysik, Technische Universit\"at Dresden, 01062 Dresden, Germany}

\date{\today}

\begin{abstract}
The Weyl semimetal NbP exhibits a very small Fermi surface consisting of two electron and two hole pockets, whose fourfold degeneracy in $k$ space is tied to the rotational symmetry of the underlying tetragonal crystal lattice.
By applying uniaxial stress, the crystal symmetry can be reduced, which successively leads to a degeneracy lifting of the Fermi-surface pockets.
This is reflected by a splitting of the Shubnikov-de Haas frequencies when the magnetic field is aligned along the $c$ axis of the tetragonal lattice.
In this study, we present the measurement of Shubnikov-de Haas oscillations of single-crystalline NbP samples under uniaxial tension, combined with state-of-the-art calculations of the electronic band structure.
Our results show qualitative agreement between calculated and experimentally determined Shubnikov-de Haas frequencies, demonstrating the robustness of the band-structure calculations upon introducing strain.
Furthermore, we predict a significant shift of the Weyl points with increasing uniaxial tension, allowing for an effective tuning to the Fermi level at only 0.8\,\% of strain along the $a$ axis.
\end{abstract}
\maketitle
\section{Introduction}
The symmetry and topology of electronic band structures in crystalline materials is inherently connected to the symmetry of the underlying crystal lattice.
A powerful experimental tool for the exploration of the symmetry dependence of electronic properties is the application of uniaxial stress, as it allows to reduce the symmetry of the crystal lattice and thus provides a rather different kind of information than application of hydrostatic pressure.
In recent years, the interest in topological properties of electronic band structures surged, which also aroused the interest in studying the uniaxial strain response of materials with remarkable topological band-structure features, using both theoretical\cite{Liu2014,Young2011,Liu2011,Battilomo2019,Lima2019,Zhao2015,Zhang2016,Winterfeld2013,Zhang2011,Teshome2019,Shao2017,Tajkov2019} and experimental \cite{Mutch2019,LiuY2014,Wang2017,Schindler2017,Hwang2014} methods.
One of those materials is the semimetallic transition-metal monopnictide NbP, which has recently stimulated numerous studies\cite{Shekhar2015,Klotz2016,Lee2015,Wu2017,Liu2016,dosReis2016,Belopolski2016,Souma2016,Zheng2016,Wang2016,Niemann2017,Gooth2017,Sergelius2016,Modic2019,Neubauer2018,Xu2017,Zheng2017,Chang2016,Sun2015}.
It exhibits a very small Fermi sea consisting of two electron and two hole pockets\cite{Lee2015,Klotz2016}, whose fourfold degeneracy in $k$ space is tied the rotational symmetry of the underlying tetragonal lattice.
Cyclotron masses in the order of 5-15\% of the free electron mass $m_0$ and large charge-carrier mobilities in the order of $10^6-10^7\,\mathrm{cm^2\,V^{-1}\,s^{-1}}$\cite{Shekhar2015} give rise to an extremely large magnetoresistance (MR) of up to $10^{6}\%$ at 1.8\,K and 9\,T in the cleanest NbP samples\cite{Shekhar2015}, superimposed by pronounced Shubnikov-de Haas (SdH) oscillations.
The small and highly degenerate pockets and the accompanied experimental access via SdH oscillations render NbP a promising platform for studying the symmetry-breaking of the electronic bands induced by uniaxial stress.
As a member of the TaAs family, NbP is predicted\cite{Huang2015,Weng2015} to exhibit 12 pairs of Weyl points, i.e., crossings of valence and conduction bands with nearly linear energy-momentum dispersion which are topologically protected by the broken inversion symmetry of the crystal lattice.
They can be classified into two four- and eightfold energy-degenerate groups, whose degeneracy is also tied to the rotational symmetry of the crystal lattice.
The existence of Weyl points near the Fermi level is thought to account for the ultrahigh mobilities observed in NbP\cite{Shekhar2015,Wang2016}.
The Fermi surface of NbP has been studied\cite{Klotz2016} theoretically by means of \emph{ab-initio} density functional theory (DFT) calculations as well as experimentally by analyzing the de Haas-van Alphen oscillations found in the magnetic-torque signal.
Measuring magnetic quantum oscillations by means of SdH experiments, also the evolution of the Fermi surface of NbP with applying hydrostatic pressure has been probed\cite{dosReis2016}.
These experiments showed a relatively stable behavior of the SdH frequencies but significant changes of their amplitudes, indicating slight deformation of the electron and hole pockets.
The effect of uniaxial stress on the electronic band structure of NbP has not yet been investigated, however, it is particularly interesting as the stress-induced breaking of the fourfold rotational symmetry is expected to lift the degeneracies of the Fermi pockets and the Weyl points.
In this paper, we report on the effect of uniaxial stress on the Fermi surface of NbP, determined via SdH oscillations in the MR of two single-crystalline NbP samples (Sample 1 and 2) subjected to direct application of uniaxial tension in a three-piezo stack apparatus (CS100, Razorbill Instruments Ltd).
For Sample 1, the magnetic field $\bm{H}$ was applied along a high-symmetry direction, resulting in a splitting of the SdH frequencies upon applying uniaxial stress due to the breaking of the fourfold rotational symmetry and thus lifting of the Fermi-pocket degeneracy.
In contrast, the SdH frequencies in Sample 2, for which $\bm{H}$ has been applied along a low-symmetry direction, did not exhibit splitting and just shifted upon increasing strain, as in that direction no crystalline symmetry is broken.
Both trends are consistent and reversible with varying $\varepsilon_{1}$, and are in good agreement with our DFT calculations.
\section{Methods}
\begin{figure}[h]
	\centering
	\includegraphics[width=8.6cm]{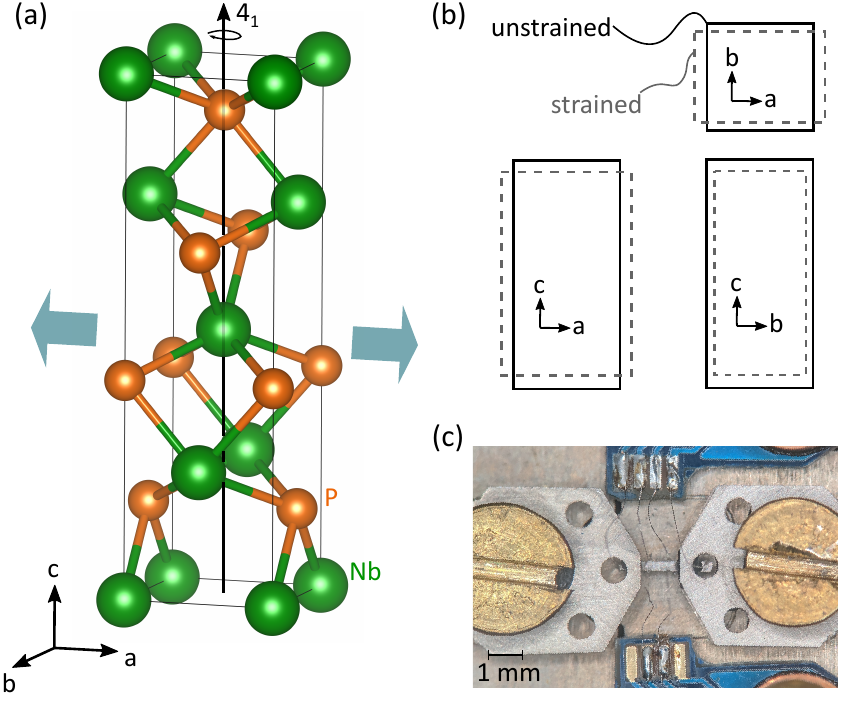}
	\caption{(a) Unit cell of the tetragonal crystal lattice of NbP. Uniaxial (tensile) stress is applied along the $a$ axis. The $4_1$ screw rotation symmetry is applicable only for the unstrained lattice. (b) Schematic view of the three lattice planes in both relaxed and strained state. When tensed along $a$, the material experiences compression along $b$ and $c$. Symmetry breaking of the lattice only occurs in the $a$-$b$ plane, where the fourfold rotational symmetry reduces to a twofold one. (c) A single-crystalline NbP sample with electrical contacts mounted on the three-piezo stack cryogenic strain cell.}
	\label{fig:crystal}
\end{figure}
NbP has a non-centrosymmetric tetragonal crystal lattice (space group $I4_1md$, No. 109) with the lattice parameters  $a=b=3.3324(2)\,\si{\angstrom}$ and $c=11.13705(7)\,\si{\angstrom}$\cite{Greenblatt1996}.
The crystal structure is shown in Fig.~\ref{fig:crystal}(a).
Single crystals of NbP were grown using chemical vapor transport reactions and oriented using X-ray diffraction.
We cut two $\sim$3\,mm-long bar-shaped samples along the $a$ axis from the same single-crystalline piece.
Sample 1 is $280(10)\,\mathrm{\mu m}$ wide and $75(5)\,\mathrm{\mu m}$ thick, with the shortest side along the $c$ axis.
Sample 2 is $310(10)\,\mathrm{\mu m}$ wide and $140(5)\,\mathrm{\mu m}$ thick, with the shortest side along the $b$ axis.
In order to determine the elastic response of NbP to applied uniaxial stress, the elastic constants have been measured on a third sample of the same batch using an ultrasound pulse-echo technique at liquid-nitrogen temperature.
For that, the main crystal planes as well as the [110]-planes were carefully polished and two lithium niobate ($\mathrm{LiNbO_3}$) transducers (Z-cut for longitudinal waves and X-cut for transverse waves) were glued to opposite parallel surfaces for excitation and detection of elastic waves.
Ultrasound propagation was measured at $27-100\,\mathrm{MHz}$ with sound-pulse durations ranging from $50-200\,\mathrm{ns}$.
For studying the SdH oscillations, electrical-transport measurements have been carried out at temperatures between $2$ and $10\,\mathrm{K}$ in a Quantum Design Physical Property Measurement system (PPMS) equipped with a 9\,T magnet (results shown only for $T=2\,\mathrm{K}$).
In all measurements, current has been applied along the $a$ axis, i.e., parallel to the uniaxial stress $\sigma_1$.
The resistivity $\rho_{xx}$ and Hall resistivity $\rho_{xy}$ have been measured in a 4-probe configuration using Stanford SR-830 lock-in amplifiers with a reference frequency of 117.23\,Hz and an excitation current of 1\,mA.
Electrical contacts were fabricated using $25\,\mathrm{\mu m}$ Pt wire and silver paint.
After characterizing the samples in the unstrained state, they were subsequently mounted to the strain cell using titanium plates and spacers with the same height as the samples, glued from top and bottom with insulating expoxy (Stycast 2850FT, Loctite) and tightened with brass screws [see Fig.~\ref{fig:crystal}(c)].
The length of the strained part of the sample is approximately 1-1.5\,mm.
The displacement is measured via a pre-calibrated capacitor, using an LCR-Meter (E4980 AL, Keysight) at 27.7\,kHz.
The peculiarities of this strain cell are extensively covered in Ref.~\onlinecite{Hicks2014}.
Notably, the size and homogeneity of the effective strain sensitively depend on a number of parameters, most importantly the sample's spring constant $k_\mathrm{s}$ as well as the thickness and stiffness of the epoxy.
The setup with fixation from both bottom and top of the sample has been reported\cite{Hicks2014} to exhibit a reasonable degree of homogeneity in the center part of the strained sample, which is where we probed the voltage.
At 2\,K, the maximum capacitance range in our strain cell was 880-945\,fF for Sample 1 and 925-950\,fF for Sample 2; in that range the displacement (few $\mathrm{\mu m}$) as a function of capacitance can be approximated as linear.
Overreading of the displacement due to device deformation is taken into account with a correction factor $(1+ G\cdot k_\mathrm{s}/k_\mathrm{\tau})^{-1}$, where $k_\mathrm{\tau}$ is the torsional stiffness of the strain cell and $G$ accounts for geometric details.
Strain losses due to the finite shear stiffness of the epoxy are accounted for with another correction factor, which is taken to be 80\% as estimated in previous works on this type of strain cell and same epoxy\cite{Steppke2017,Jo2019}.
During the mounting procedure, heat curing of the epoxy and thus thermal expansion of the piezoelectric stacks and other parts of the chassis lead to a deviation of the sample from the zero-strain state with respect to the calibration when cooled back to room temperature.
We account for this by subtracting a constant offset, such that a linear extrapolation of the SdH frequencies under strain intersects with the SdH frequency in the unstrained state.
This yielded a moderate compression of $-0.17\,\%$ for Sample 1 and $-0.1\,\%$ for Sample 2 at room temperature without having any voltage applied to the piezoelectric stacks.
For our \textit{ab-initio} investigations of the electronic band structure, we employ DFT as implemented in the \textsc{VASP} package~\cite{kresse1996}.
Here, plane waves with pseudopotentials are used as a basis set and the generalized-gradient approximation (GGA)~\cite{perdew1996} is utilized for the description of the exchange-correlation potential.
For calculating fine meshes in the Brillouin zone we create a tight-binding Hamiltonian from Wannier functions using the \textsc{Wannier90} package~\cite{Mostofi2008} with initial projections to the $d$ orbitals of Nb and to the $p$ orbitals of P.
For the evaluation of the extremal orbits a $k$ mesh with a resolution better than 0.004 $\si{\angstrom}^{-1}$ was used.
\section{Elastic properties}
To determine the elastic response for a given applied uniaxial stress and to be able to estimate the effectively applied strain in the samples, the elastic stiffness tensor $C_{ij}$ of NbP needs to be considered.
Using Voigt notation ($xx\rightarrow 1$, $yy\rightarrow 2$, $zz\rightarrow 3$, $xz\rightarrow 4$, $yz\rightarrow 5$, $xy\rightarrow 6$), the sample's strain components $\varepsilon_j$ are related to the applied stress components via $\sigma_i=\sum_j C_{ij}\varepsilon_j$.
For a tetragonal crystal $C_{ij}$ holds
\begin{equation}
C = 
\begin{bmatrix}
C_{11} & C_{12} & C_{13} & 0 & 0 & 0\\
C_{12} & C_{11} & C_{13} & 0 & 0 & 0\\
C_{13} & C_{13} & C_{33} & 0 & 0 & 0\\
0 & 0 & 0 & C_{44} & 0 & 0\\
0 & 0 & 0 & 0 & C_{44} & 0\\
0 & 0 & 0 & 0 & 0 & C_{66}\\
\end{bmatrix}.
\end{equation}
We determined all elastic constants except for $C_{13}$ via measurement of the speed of sound $v=\sqrt{C_\mathrm{eff}/\rho}$ for longitudinal and transverse modes along the main axes as well as along [110], $\rho$ being the mass density ($\rho=6.52\,\mathrm{g\,cm^{-3}}$ for NbP\cite{Greenblatt1996}).
The results are shown in Tab.~\ref{tab:ElasticConstants} and show good agreement with the values calculated from DFT\cite{Liu2017}.
Applying uniaxial tension along $a$ ($\sigma_{1}>0$, $\sigma_i$ = 0 for $i\neq 1$) results in strain $\varepsilon_1=\sigma_{1}/E$ along the same axis, where the Young's modulus $E$ in that configuration holds
\begin{equation}
	E=\frac{1}{S_{11}}=\frac{(C_{11}-C_{12})[(C_{11}+C_{12})C_{33}-2C_{13}^2]}{C_{11}C_{33}-C_{13}^2},
\end{equation}
$S_{11}$ being the first component of the elastic compliance tensor $S_{ji}$.
Given the good agreement between experimentally determined and theoretical elastic constants, we used the calculated value\cite{Liu2017} $C_{13}=130.8\,\mathrm{GPa}$ to estimate the Young's modulus, yielding $E=210(60)\,\mathrm{GPa}$ ($E=251\,\mathrm{GPa}$ using only the calculated values).
With $k_\mathrm{s}=EA/l$, $A$ being the sample cross section area and $l$ its length, the sample spring constant yields $(4.4 \pm 1.4)\times 10^6\,\mathrm{Nm^{-1}}$ for Sample 1 and $(6.3 \pm 1.8)\times 10^6\,\mathrm{Nm^{-1}}$ for Sample 2.
This results in an about $15\%$ and $25\%$ too large displacement value due to cell deformation in Sample 1 and Sample 2, respectively.
Naturally, tension along $a$ results in compression along $b$ with the Poisson ratio
\begin{equation}
\nu_{21}=-\frac{\varepsilon_2}{\varepsilon_1}=-\frac{S_{12}}{S_{11}}=\frac{C_{12}C_{33}-C_{13}^2}{C_{11}C_{33}-C_{13}^2}
\end{equation}
and along $c$, respectively, with 
\begin{equation}
\nu_{31}=-\frac{\varepsilon_3}{\varepsilon_1}=-\frac{S_{13}}{S_{11}}=\frac{(C_{11}-C_{12})C_{13}}{C_{11}C_{33}-C_{13}^2}.
\end{equation}
This yields $\nu_{21}=0.52(16)$ and $\nu_{31}=0.23(11)$ ($\nu_{21}=0.36$ and $\nu_{31}=0.29$ from calculated values).
In our DFT calculations, the latter were used.
The effect of uniaxial stress $\sigma_1$ on the crystal symmetry is illustrated in Fig.~\ref{fig:crystal}(b): the fourfold rotational symmetry, more precisely the screw operation $4_1$ parallel to the $c$ axis of the tetragonal lattice, is broken and the crystal symmetry reduces to an orthorhombic lattice.
In our setup, only the displacement $\Delta l$ along the $a$ axis is measured, allowing for the determination of $\varepsilon_1=\Delta l/l$.
\begin{table}[ht!]
	\caption{Elastic constants of NbP determined from ultrasound pulse-echo measurements.}
	\label{tab:ElasticConstants}
	\begin{ruledtabular}
		\begin{tabular}{c|c|c|c}
			Mode & $v (\mathrm{km/s})$ & $C_{ij} \mathrm{(exp.)} (\mathrm{GPa})$ & $C_{ij} \mathrm{(theor.)} (\mathrm{GPa})$\footnote{The theoretical values are taken from Ref.~\onlinecite{Liu2017}.} \\ \hline
			$C_{11}$ & 7.33(14) & 350(13) & 349.2 \\
			$C_{33}$ & 6.4(5) & 270(40) & 285.9 \\
			$C_{44}$ & 4.09(5) & 109(3) & 111.9 \\
			$C_{66}$ & 5.53(8) & 200(6) & 199.7 \\
			$\frac{1}{2}(C_{11}-C_{12})$ & 3.2(2) & 210(40) ($C_{12}$) & 164.7 ($C_{12}$) \\ 
		\end{tabular}
	\end{ruledtabular}
\end{table}
\section{Results}
\begin{figure*}[ht!]
	\centering
	\includegraphics[width=17.2cm]{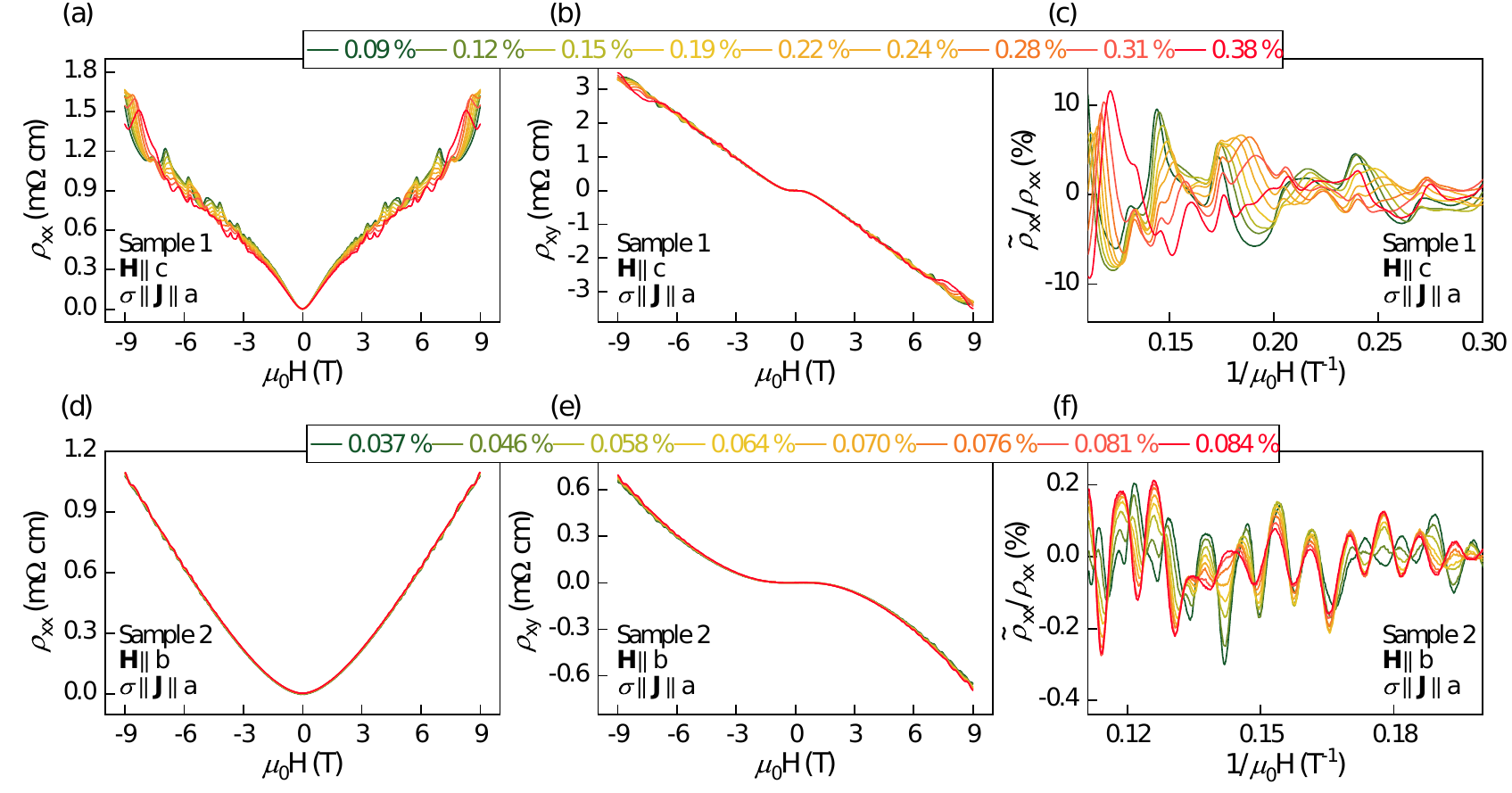}
	\caption{Electrical magnetotransport data of two single-crystalline NbP samples for different tensile strains $\varepsilon_{1}>0$. Magnetic field $\bm{H$} is applied along $c$ for Sample 1 and along $b$ for Sample 2. Uniaxial stress is applied along $a$ for both samples. (a),(d) Resistivity $\rho_{xx}$ versus magnetic field strength $H$. (b),(e) Hall resistivity $\rho_{xy}$ versus $H$. (c),(f) Shubnikov-de Haas oscillations extracted from $\rho_{xx}(H)$ versus $1/H$.}
	\label{fig:MR}
\end{figure*}
\begin{figure}[h]
	\centering
	\includegraphics[width=8.6cm]{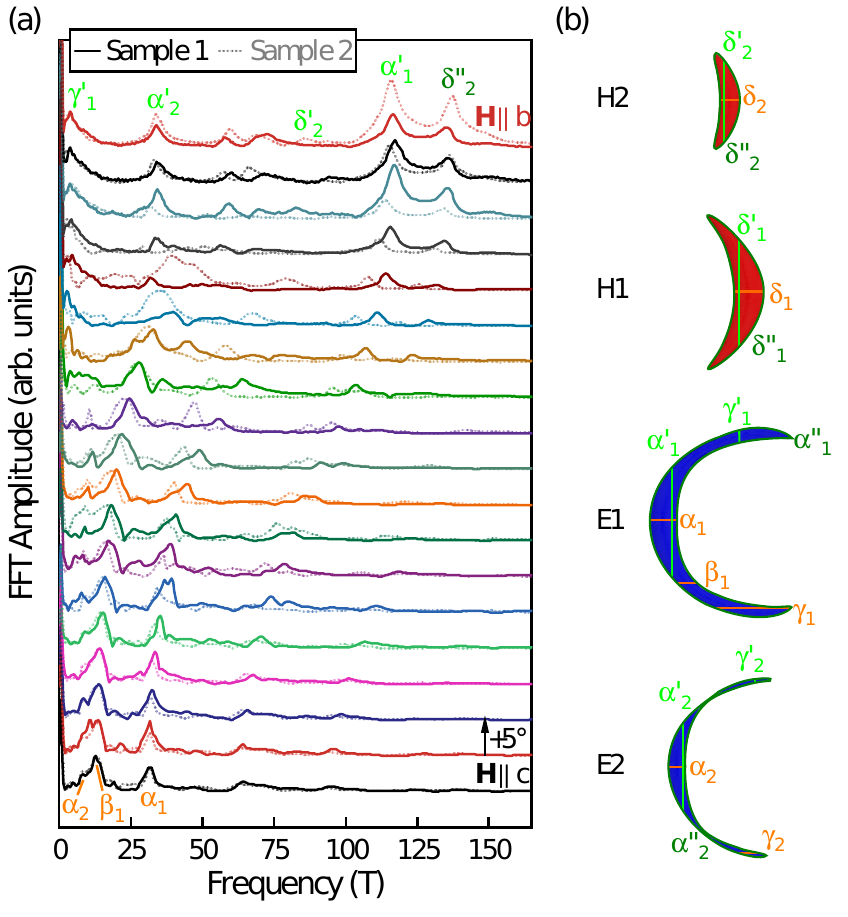}
	\caption{(a) Angular variation of the Fourier spectrum of the SdH oscillations for both samples before mounting to the strain cell. (b) Plot of the two electron (blue) and two hole pockets (red) and the extremal orbits for fields along $b$ and $c$.}
	\label{fig:Rotation}
\end{figure}
\begin{figure}[h]
	\centering
	\includegraphics[width=8.6cm]{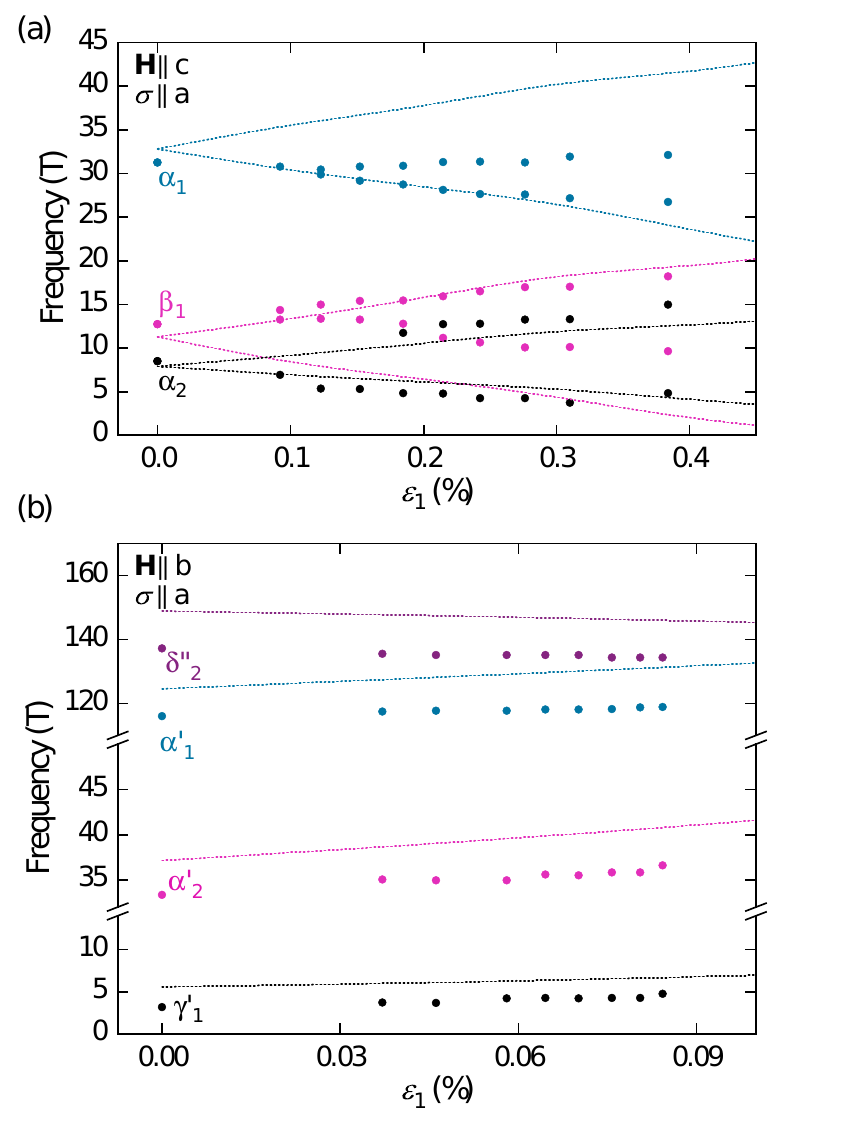}
	\caption{Comparison of experimental (dots) and theoretical SdH frequencies (dashed lines) for Sample 1 (a) and Sample 2 (b).}
	\label{fig:Results}
\end{figure}
\begin{figure*}[ht!]
	\centering
	\includegraphics[width=17.2cm]{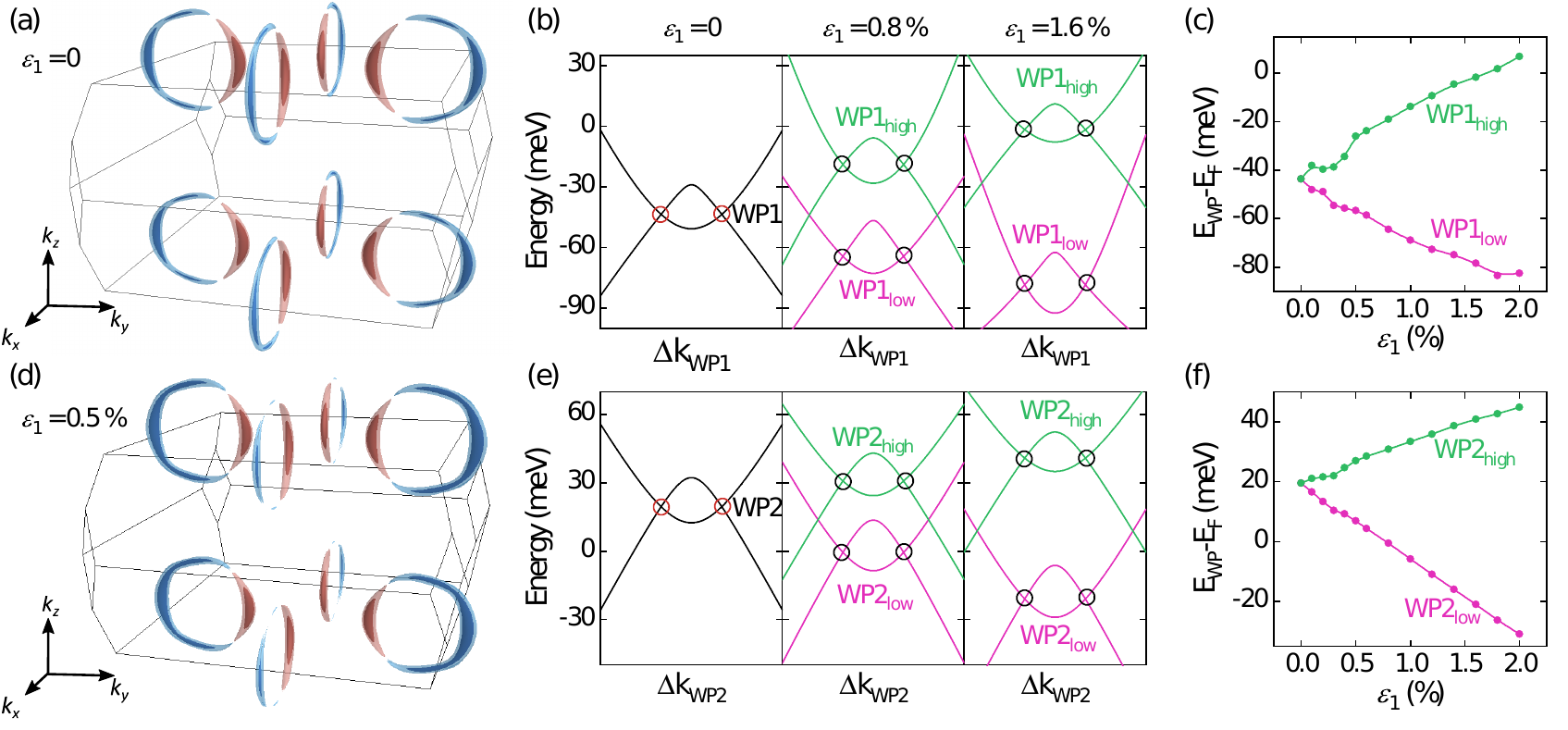}
	\caption{DFT Results of NbP under uniaxial tension. (a),(d) Fermi surface of NbP in the first Brillouin zone for (a) the unstrained lattice and (d) $\varepsilon_{1}=0.5\%$. (b),(e) Evolution of Weyl point pairs WP1 and WP2 under uniaxial tension. WP1 is fourfold degenerate in $k$ space and WP2 eightfold degenerate, respectively. Upon application of uniaxial stress, they split into twofold and fourfold degenerate ones, respectively. The horizontal axis is displayed in arbitrary units of the $k$ vector connecting the corresponding pair, however, the split pairs denoted ``high'' and ``low'' are at different positions in the Brillouin zone. (c),(f) Energy distance of the Weyl points from the Fermi level. Upon increasing tension, $\mathrm{WP1_{high}}$ and $\mathrm{WP2_{low}}$ can be tuned to the Fermi level at moderate strains of 1.6\,\% and 0.8\,\%, respectively.}
	\label{fig:Brillouinzone}
\end{figure*}

The MR $\rho_{xx}(H)$ and Hall resistivity $\rho_{xy}(H)$ of both samples are shown at different (tensile) strains $\varepsilon_{1}>0$ in Fig.~\ref{fig:MR}, whereby $\bm{H}$ has been applied along $c$ for Sample 1 and along $b$ for Sample 2.
Both samples exhibit a large MR characteristic for NbP\cite{Shekhar2015}, superimposed by SdH oscillations.
In agreement with previous reports, the amplitude of the SdH oscillations strongly varies depending on the $\bm{H}$ direction and is most pronounced for $\bm{H}$ along $c$, which is also apparent comparing the MR in Sample 1 and Sample 2.
Quantum oscillations are also superimposed on $\rho_{xy}(H)$ indicating an oscillating number of charge carriers and a somewhat fixed Fermi level.
Likewise, the oscillations in $\rho_{xy}(H)$ are more pronounced for $\bm{H}\parallel c$ [Figs.~\ref{fig:MR}(b) and \ref{fig:MR}(e)].
A continuous evolution of the quantum oscillations with increasing strain is observed, better visible in the oscillatory part $\tilde{\rho}_{xx}/\rho_{xx}$ versus $1/H$ [Figs.~\ref{fig:MR}(c) and \ref{fig:MR}(f)], which has been extracted from the MR by subtracting a smooth polynomial of third order.
The strain response of $\rho_{xx}$ and $\rho_{xy}$ was completely elastic, i.e., no hysteresis was observed upon increasing or decreasing the stress and all curves could be reproduced by applying the same stress again.
This provides an ideal playground for in-situ manipulation of the band structure, without introducing additional disorder or facing the irreversibility of high-pressure experiments.
Notably, both MR and Hall resistivity do not exhibit significant changes apart from the superimposed oscillations, which shows that the deformation of the Fermi surface does not have a strong effect on the general charge-carrier properties, e.g., density and mobility, but rather leads to a redistribution of the carriers in $k$ space.
This is in contrast to the effect of hydrostatic pressure on NbP\cite{dosReis2016}, where both MR and Hall resistivity change considerably.
By Fourier transformation of $\tilde{\rho}_{xx}/\rho_{xx}(H^{-1})$, a characteristic spectrum is revealed.
Via the Onsager relation\cite{Shoenberg} $F=(\Phi_0/2\pi^2)A_\mathrm{ext}$, each SdH frequency $F$ is related to an extremal orbit of the Fermi surface plane perpendicular to the applied $\bm{H}$, where $A_\mathrm{ext}$ is the area enclosed by this orbit and $\Phi_0$ is the magnetic flux quantum.
The SdH spectra of the two samples in the unstrained state are shown in Fig.~\ref{fig:Rotation}(a), for $\bm{H}$ rotated from the $c$ to the $b$ axis.
Noticeable differences in the intermediate range between 20 and $70^\circ$ occur, presumably due to a bigger sensitivity to misalignments in that range.
However, for $\bm{H}$ along $b$ and $c$ the spectrum of the two samples is in good agreement and is roughly stable upon misalignments of $\pm 5^\circ$.
This is expected, as both samples are cut from the same piece and thus the Fermi level should be nearly identical.
The $\bm{H}$ direction in our strain setup is fixed, thus, the matching of the SdH spectra is a necessary check to justify the comparison of the two samples.
The Fermi surface of NbP is shown in Fig.~\ref{fig:Brillouinzone}(a).
There are a number of extremal orbits [schematically shown in Fig.~\ref{fig:Rotation}(b)], which we labeled as in Ref.~\onlinecite{Klotz2016}, resulting in a complex spectrum.
The two hole pockets, H1 and H2, exhibit only maximum orbits.
For $\bm{H}$ along $c$, the orbit is termed as $\delta_1$ ($\delta_2$).
For $\bm{H}$ along $b$ or $a$, the fourfold rotational symmetry yields two maximum orbits per hole pocket, termed as $\delta'_1$ and $\delta''_1$ ($\delta'_2$ and $\delta''_2$).
The two electron pockets, E1 and E2, exhibit a number of both maximum and minimum orbits.
For $\bm{H}$ along $c$, E1 (E2) exhibits two maxima, $\alpha_1$ ($\alpha_2$) and $\gamma_1$ ($\gamma_2$), and a minimum orbit $\beta_1$ [$\beta_2$ yields a SdH frequency smaller than 1\,T and is not shown in Fig.~\ref{fig:Rotation}(b)].
For $\bm{H}$ along $b$ or $a$, there are three possible orbits for E1 (E2), termed $\alpha'_1$ ($\alpha'_2$), $\alpha''_1$ ($\alpha''_2$), and $\gamma'_1$ ($\gamma'_2$).
Variation of the Fermi level to 11\,meV above the charge-neutral point resulted in the best matching between calculated and experimental SdH frequencies.
However, we could not attribute each orbit to a peak in the experimental spectra.
Also, we observe experimental peaks in a range where DFT yields no fundamental frequencies, some of which can be attributed to higher harmonics.
Some orbits yield very similar SdH frequencies, as for example $\alpha_1$ (32.8\,T) and $\gamma_1$ (31.1\,T), which falls below our experimental resolution.
In case of two calculated frequencies lying close to a peak in the experimental spectrum, we chose the orbit with the highest congruency of the calculated and experimental $\mathrm{d}F/\mathrm{d}\varepsilon_{1}$.
This procedure led to the same assignments of experimental and calculated values as in Refs.~\onlinecite{Klotz2016,dosReis2016}.
Further challenges of disentangling the rich spectrum of NbP are thoroughly covered in Ref.~\onlinecite{Klotz2016}. 
In the following, we focus on Fourier peaks which are distinct and could be attributed to a theoretical orbit with good agreement, marked in Fig.~\ref{fig:Rotation}(a) and summarized in Table~\ref{tab:Data}.
\begin{table}[h!]
	\caption{Experimental and calculated values for the SdH frequencies of selected orbits and their strain derivatives.}
	\label{tab:Data}
	\begin{ruledtabular}
		\begin{tabular}{c|cc|cc}
			Orbit & Theory &  & Experiment & \\
			\hline
			\multicolumn{5}{c}{$\bm{H}\parallel c$ (Sample 1)} \\
			\hline
			& $F(\mathrm{T})$ & $\mathrm{d}F/\mathrm{d}\varepsilon_{1}(\mathrm{T}/\%)$ & $F(\mathrm{T})$ & $\mathrm{d}F/\mathrm{d}\varepsilon_{1}(\mathrm{T}/\%)$ \\
			\hline
			\multirow{2}{*}{$\alpha_2$} & \multirow{2}{*}{7.92} & 13.81 & \multirow{2}{*}{8.5} & 14(2)\\
			&  & -8.37 &  & -7(2) \\
			\multirow{2}{*}{$\beta_1$} & \multirow{2}{*}{11.3} & 24.08 & \multirow{2}{*}{12.7} & 13(1) \\
			&  & -22.1 &  & -15(2) \\
			\multirow{2}{*}{$\alpha_1$} & \multirow{2}{*}{32.8} & 25.12 & \multirow{2}{*}{31.3} & 7(1) \\
			&  & -20.2 &  & -14(2)\\
			\hline
			\multicolumn{5}{c}{$\bm{H}\parallel b$ (Sample 2)} \\
			\hline
			& $F(\mathrm{T})$ & $\mathrm{d}F/\mathrm{d}\varepsilon_{1}(\mathrm{T}/\%)$ & $F(\mathrm{T})$ & $\mathrm{d}F/\mathrm{d}\varepsilon_{1}(\mathrm{T}/\%)$ \\
			\hline
			$\gamma'_1$ & 5.6 & 11 & 3.2 & 18(4) \\
			$\alpha'_2$ & 37.17 & 40.3 & 33.4 & 29(7) \\
			$\alpha'_1$ & 124.56 & 77.6 & 116 & 29(4) \\
			$\delta''_2$ & 148.89 & -30.6 & 137 & -26(5) \\
		\end{tabular}
	\end{ruledtabular}
\end{table}
\begin{figure}[ht]
	\centering
	\includegraphics[width=8.6cm]{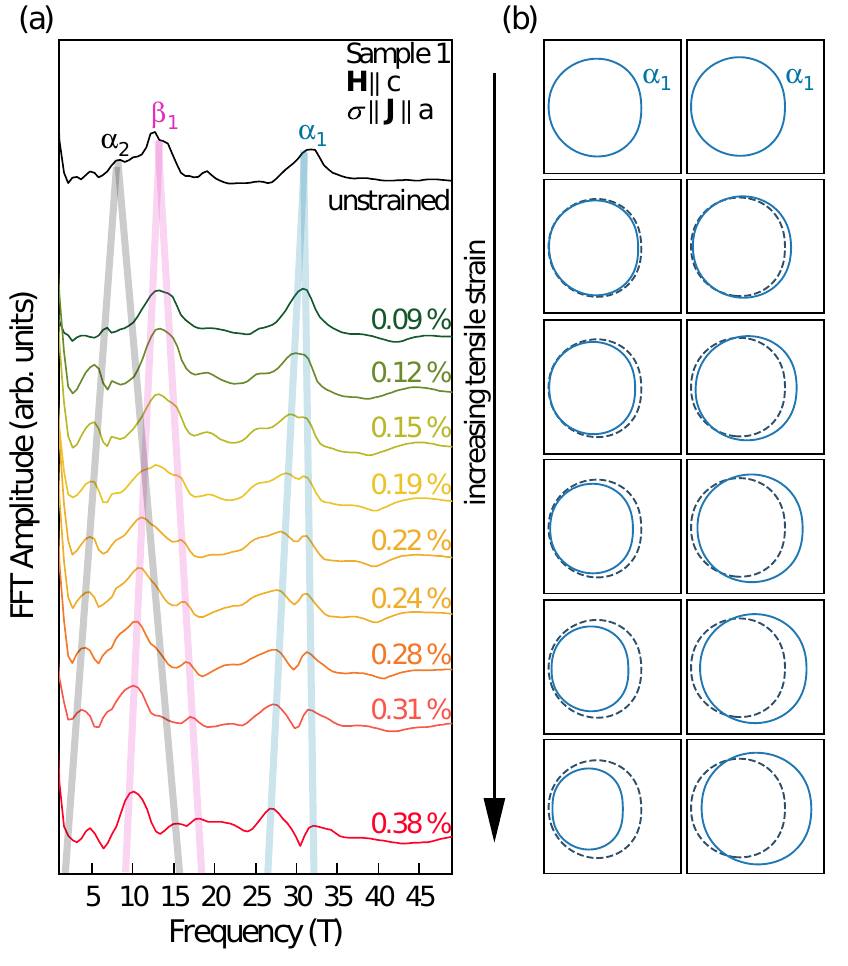}
	\caption{(a) SdH spectra of Sample 1 ($\bm{H}\parallel c$) for the different tensile strains as well as the unstrained case.  (b) Splitting of the $\alpha_1$ orbit upon increasing strain (0, 0.1, 0.2, 0.3, 0.4, 0.5\%) along $a$ from DFT. The left (right) column depicts the $\alpha_1$ orbit with decreasing (increasing) SdH frequency.}
	\label{fig:Strainc}
\end{figure}
\begin{figure}[ht]
	\centering
	\includegraphics[width=8.6cm]{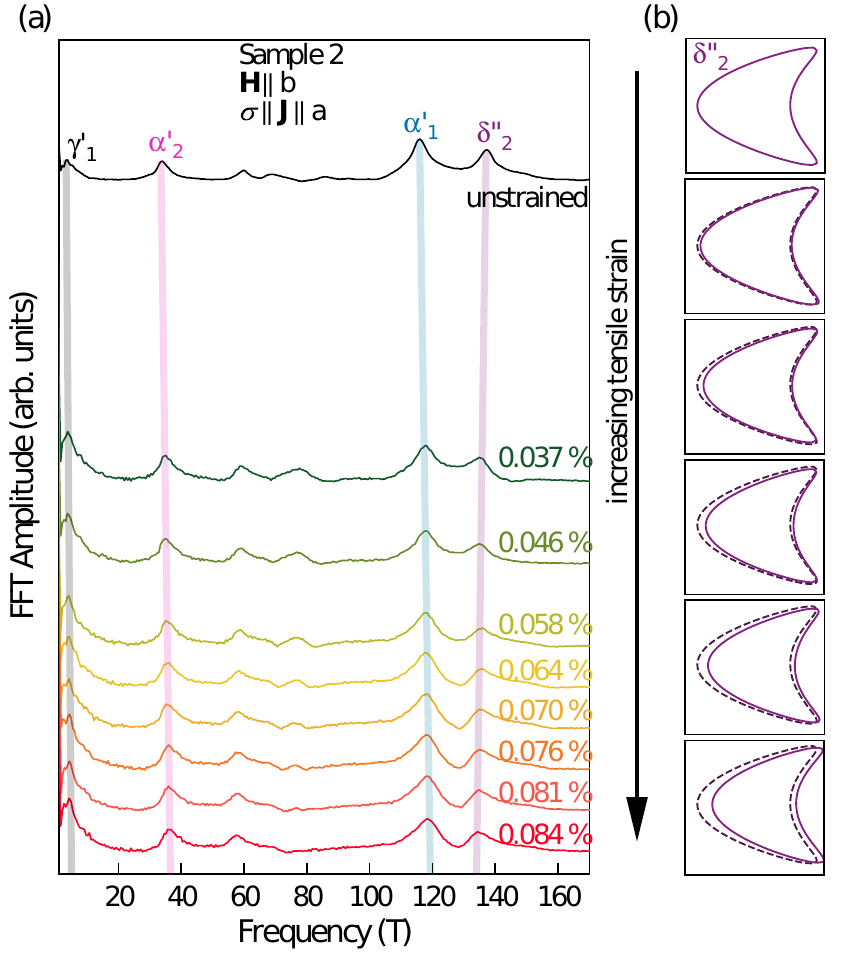}
	\caption{(a) Fourier spectrum of Sample 2 ($\bm{H}\parallel b$) for the different tensile strains as well as the unstrained case. (b) The evolution of the $\delta''_2$ orbit with increasing tensile strain (0, 0.1, 0.2, 0.3, 0.4, 0.5\%) is illustrated (DFT results).}
	\label{fig:Strainb}
\end{figure}
The Fourier spectra for $\bm{H}\parallel c$ (Sample 1) for different strains are shown in Fig.~\ref{fig:Strainc}(a).
A clear splitting of the $\alpha_1$ and $\beta_1$ peaks into an increasing and a decreasing branch upon increasing $\varepsilon_1$ is observed.
The evolution of $\alpha_2$ is less pronounced, however, signatures of the $\alpha_2$ peak found for some of the strain values are indicative of a splitting as well.
As the increasing branch of $\alpha_2$ is crossing the decreasing branch of $\beta_1$, and the amplitude of $\beta_1$ surpasses that of $\alpha_2$, the increasing branch of the $\alpha_2$ peak cannot be fully traced.
The splitting of the peaks is consistent with our DFT calculations and is a strong evidence for the lifting of the degeneracy of the Fermi pockets via breaking the fourfold rotational symmetry of the tetragonal crystal lattice.
As illustrated in Figs.~\ref{fig:Strainc}(b), each fourfold degenerate orbit splits into two shrinking and two expanding orbits, resulting in two branches of SdH frequencies [see also Fig.~\ref{fig:Brillouinzone}(d)].
In contrast, the Fourier peaks for $\bm{H}\parallel b$ (Sample 2) do not exhibit a splitting or broadening and just shift upon increasing strain [Fig.~\ref{fig:Strainb}(a)].
This qualitative behavior is congruent with our DFT calculations and serves as a control experiment, confirming that the splitting of the SdH frequencies is tied to the orientation of $\bm{H}$ along the $4_1$ screw axis of the crystal lattice.
The experimental and calculated SdH frequencies for the attributed orbits are plotted versus $\varepsilon_{1}$ in Fig.~\ref{fig:Results}.
All calculated trends are qualitatively reproduced by the experimental values, demonstrating that the strain evolution of the Fermi surface of NbP is well described by DFT.
We note, however, that the absolute values of the majority of experimentally determined frequencies differ from the calculated values by $\sim$10\%. 
This deviation is of similar magnitude as in the previous reports comparing quantum oscillations and DFT results for NbP\cite{Klotz2016,dosReis2016}.
The derivatives $\mathrm{d}F/\mathrm{d}\varepsilon_{1}$ are extracted by linear fits of $F(\varepsilon_{1})$ and are given next to the DFT results in Table~\ref{tab:Data}.
The sign and order of magnitude of the $\mathrm{d}F/\mathrm{d}\varepsilon_{1}$ values are in good agreement between theory and experiment for most orbits, whereas matching in terms of absolute values for some orbits, e.g., $\alpha_1$ and $\alpha'_1$ differs by more than 100\%.
Erroneous extraction of the experimental values might be caused by the smearing of the peaks due to the field-dependent damping of the SdH amplitude.
\begin{figure}[h]
	\centering
	\includegraphics[width=8.6cm]{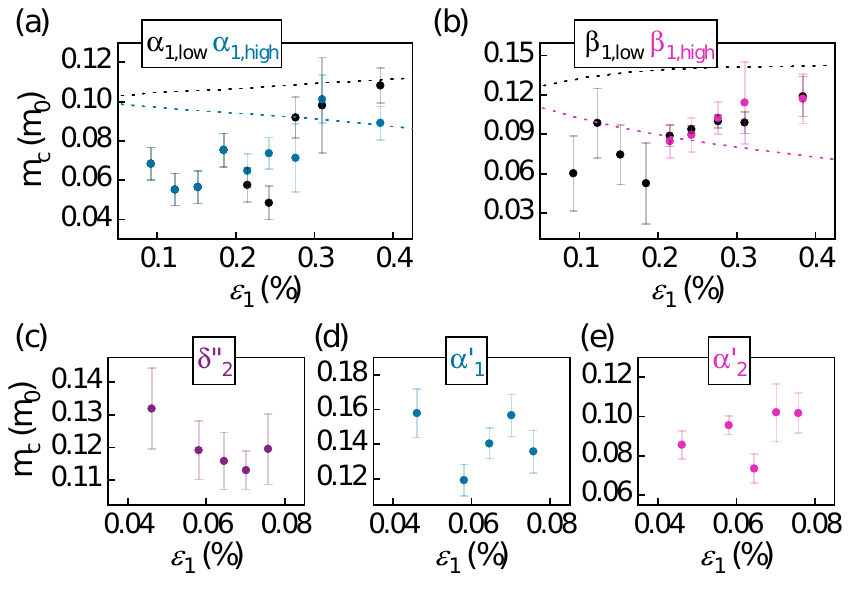}
	\caption{Strain evolution of the cyclotron masses extracted from the $T$ dependence of the Fourier peaks for different orbits. Dotted lines represent the calculated values.}
	\label{fig:Mass}
\end{figure}
The cyclotron masses $m_\mathrm{c}=\hbar^2/2\pi\cdot\partial A/\partial E$ of the corresponding orbits of area $A=A_\mathrm{ext}$ at the Fermi energy $E_\mathrm{F}$ are extracted from the $T$-dependent damping of the Fourier peaks via\cite{Shoenberg}
\begin{equation}
R_\mathrm{T}=\frac{\lambda (T)}{\sinh[\lambda(T)]},\mathrm{with}~\lambda(T)=2\pi^2 \frac{m_\mathrm{c}}{eH}\frac{k_\mathrm{B}T}{\hbar},
\label{eq:tempdamping}
\end{equation}
where $R_\mathrm{T}$ is the amplitude factor, $\hbar$ the reduced Planck constant, $e$ the electron charge, and $k_\mathrm{B}$ the Boltzmann constant.
$1/H$ is taken as the average $(1/H)_\mathrm{avg}$ of the Fourier window $\Delta(1/H)$.
We extracted $m_\mathrm{c}$ for several orbits (Fig.~\ref{fig:Mass}) from the Fourier spectra at $T=2-10\,\mathrm{K}$.
Close to zero strain, the experimental values for $\alpha_1$ and $\beta_1$ are in good agreement with those reported in Ref.~\onlinecite{Klotz2016}, and similarly deviate from the higher calculated values.
Cyclotron masses extracted from Fourier spectra, especially with frequencies below $100\,\mathrm{T}$, are usually underestimated by up to 50-60\,\%\cite{Audourd2018,Mercure2008} due to the strong field dependence of the SdH amplitude.
The extracted mass heavily depends on $\Delta(1/H)$ and $(1/H)_\mathrm{avg}$, whereas decreasing $\Delta(1/H)$ generally yields $m_\mathrm{c}$ values closer to the real one.
However, as the sampling interval of the discrete Fourier spectrum is $\left(\Delta(1/H)\right)^{-1}$, decreasing $\Delta(1/H)$ drastically reduces the resolution and makes identifying distinct peaks for $\bm{H}\parallel c$ (Sample 1) impossible.
Additional complication arises from the overlapping of the Fourier peaks.
Therefore, the large error bars in Fig.~\ref{fig:Mass} are estimated by extracting $m_\mathrm{c}$ for different $\Delta(1/H)$ and $(1/H)_\mathrm{avg}$ as well as fitting Eq.~(\ref{eq:tempdamping}) to Fourier spectra of simulated data with similar frequencies and cyclotron masses.
Notably, envelope extraction via inverse Fourier transformation of a frequency window (with rounded corners\cite{Mercure2008}) enclosing a Fourier peak of the simulated data also did not result in reliable determination of $m_\mathrm{c}$.
Due to the large uncertainty, we were not able to resolve the splitting of $m_\mathrm{c}$ predicted by DFT; the extracted values can only be considered as a rough estimate.
The cyclotron masses for $\bm{H}\parallel b$ (Sample 2) are presumably slightly more reliable, as the peaks are more distinguishable and a smaller $\Delta(1/H)$ window could be used. 
The overall qualitative (and in some parts quantitative) agreement of the DFT calculations with the SdH data might as well allow us to draw conclusions about the Weyl points in NbP under uniaxial stress.
The 12 pairs of Weyl points in NbP fall in two groups: four pairs in the $k_z=0$ plane (WP1) and eight pairs in the $k_z\approx\pm\pi/c$ plane (WP2).
Upon increasing $\varepsilon_1$, the four (eight) WP1 (WP2) pairs split into two types of two (four) pairs each, labelled ``high'' and ``low'' [see Figs.~\ref{fig:Brillouinzone}(b),(e)], shifting up and down in energy, respectively.
Notably, the energy shift is quite significant, with tens of meV (or hundreds of K in terms of $E/k_\mathrm{B}$) at a few percent of strain [Fig.~\ref{fig:Brillouinzone}(c),(f)].
Predicted by our calculations, the $\mathrm{WP1_{high}}$ pairs should be situated at the Fermi level at $\varepsilon_{1}\approx 1.6\,\%$, and $\mathrm{WP2_{low}}$ pairs at $\varepsilon_{1}\approx 0.8\,\%$, respectively.
In contrast to the application of hydrostatic pressure\cite{dosReis2016}, where several GPa are required to tune WP2 a few meV closer to the Fermi level and WP1 does not exhibit any notable shift, uniaxial stress might allow for effective tuning of the Weyl points without having to deal with additional disorder as is the case with chemical doping.
As epitaxial thin films of NbP have already been grown\cite{Bedoya2020}, realization of strained thin films might render a promising opportunity towards ``Weyl-tronic'' devices.
\section{Summary}
In summary, we studied the response of SdH oscillations in NbP to direct application of uniaxial stress and combined the experimental results with DFT calculations.
By applying stress along one of the axes of the square $ab$ plane of the tetragonal crystal lattice, the fourfold rotational symmetry is broken.
This is reflected by a splitting of the peaks in the Fourier spectra of the SdH oscillations for $\bm{H}$ along the $c$ axis.
When $\bm{H}$ is applied along one of the equal axes, the symmetry of the orbits on the Fermi surface remains unchanged, thus only shifting of the Fourier peaks is observed.
Our observations are qualitatively congruent with the DFT calculations, demonstrating their robustness for NbP in the strained state.
Furthermore, we predict a significant energy shifting of the Weyl points, whereas increasing tensile strain of one percent leads to a shift of tens of meV.
This might be an effective way to tune the Weyl points to the Fermi level in order to realize Weyl-based devices.
\begin{acknowledgments}
	C.~S. thanks R. Koban and H. Borrmann for providing technical support as well as D. Gorbunov and S. Zherlitsyn for assistance with the ultrasound pulse-echo measurements.
	C.~S. acknowledges financial support by the International Max Planck Research School for Chemistry and Physics of Quantum Materials (IMPRS-CPQM).
	J.~G. acknowledges support from the European Union's Horizon 2020 research and innovation program under Grant Agreement ID 829044 ``SCHINES''.
	The work was supported by the Deutsche Forschungsgemeinschaft (DFG) through SFB 1143 and the W\"urzburg-Dresden Cluster of Excellence on Complexity and Topology in Quantum Matter--$ct.qmat$ (EXC 2147, Project No. 390858490), by Hochfeld-Magnetlabor Dresden (HLD) at HZDR, member of the European Magnetic Field Laboratory (EMFL), and by the ERC Advanced Grant No. 742068 ``TOPMAT''.
\end{acknowledgments}
%
%
%
%

%

\begin{thebibliography}{10}
 	\expandafter\ifx\csname url\endcsname\relax
 	\def\url#1{\texttt{#1}}\fi
 	\expandafter\ifx\csname urlprefix\endcsname\relax\def\urlprefix{URL }\fi
 	\providecommand{\bibinfo}[2]{#2}
 	\providecommand{\eprint}[2][]{\url{#2}}
 	
 	\bibitem{Liu2014}
 	\bibinfo{author}{J.~Liu}, \bibinfo{author}{Y.~Xu}, \bibinfo{author}{J.~Wu}, \bibinfo{author}{B.-L.~Gu}, \bibinfo{author}{S.~B.~Zhang} \& \bibinfo{author}{W.~Duan},
 	\newblock \bibinfo{title}{{Manipulating topological phase transition by strain}},
 	\newblock \emph{\bibinfo{journal}{Acta Cryst. C}} \textbf{\bibinfo{volume}{70}},
 	\bibinfo{pages}{118} (\bibinfo{year}{2014}).
 	
 	\bibitem{Young2011}
 	\bibinfo{author}{S.~M.~Young}, \bibinfo{author}{S.~Chowdhury}, \bibinfo{author}{E.~J.~Walter}, \bibinfo{author}{E.~J.~Mele}, \bibinfo{author}{C.~L.~Kane} \& \bibinfo{author}{A.~M.~Rappe},
 	\newblock \bibinfo{title}{{Theoretical investigation of the evolution of the topological phase of $\mathrm{Bi_2Se_3}$ under mechanical strain}},
 	\newblock \emph{\bibinfo{journal}{Phys. Rev. B}} \textbf{\bibinfo{volume}{84}},
 	\bibinfo{pages}{085106} (\bibinfo{year}{2011}).
 	
 	\bibitem{Liu2011}
 	\bibinfo{author}{W.~Liu}, \bibinfo{author}{X.~Peng}, \bibinfo{author}{C.~Tang}, \bibinfo{author}{L.~Sun}, \bibinfo{author}{K.~Zhang} \& \bibinfo{author}{J.~Zhong},
 	\newblock \bibinfo{title}{{Anisotropic interactions and strain-induced topological phase transition in $\mathrm{Sb_2Se_3}$ and $\mathrm{Bi_2Se_3}$}},
 	\newblock \emph{\bibinfo{journal}{Phys. Rev. B}} \textbf{\bibinfo{volume}{84}},
 	\bibinfo{pages}{245105} (\bibinfo{year}{2011}).
 	 	
 	\bibitem{Battilomo2019}
 	\bibinfo{author}{R.~Battilomo}, \bibinfo{author}{N.~Scopigno} \& \bibinfo{author}{C.~Ortix},
 	\newblock \bibinfo{title}{{Tuning topology in thin films of topological insulators by strain gradients}},
 	\newblock \emph{\bibinfo{journal}{Phys. Rev. B}} \textbf{\bibinfo{volume}{100}},
 	\bibinfo{pages}{115131} (\bibinfo{year}{2019}).
 	
 	 \bibitem{Lima2019}
 	\bibinfo{author}{E.~N.~Lima}, \bibinfo{author}{T.~M.~Schmidt} \& \bibinfo{author}{R.~W.~Nunes},
 	\newblock \bibinfo{title}{{Structural and topological phase transitions induced by strain in two-dimensional bismuth}},
 	\newblock \emph{\bibinfo{journal}{J. Phys.: Condens. Matter}} \textbf{\bibinfo{volume}{31}},
 	\bibinfo{pages}{475001} (\bibinfo{year}{2019}).
 	 	
 	\bibitem{Zhao2015}
 	\bibinfo{author}{M.~Zhao}, \bibinfo{author}{X.~Zhang} \& \bibinfo{author}{L.~Li},
 	\newblock \bibinfo{title}{{Strain-driven band inversion and topological aspects in Antimonene}},
 	\newblock \emph{\bibinfo{journal}{Sci. Rep.}} \textbf{\bibinfo{volume}{5}},
 	\bibinfo{pages}{16108} (\bibinfo{year}{2015}).
 	
 	 \bibitem{Zhang2016}
 	\bibinfo{author}{S.~Zhang}, \bibinfo{author}{M.~Xie}, \bibinfo{author}{B.~Cai}, \bibinfo{author}{H.~Zhang}, \bibinfo{author}{Y.~Ma}, \bibinfo{author}{Z.~Chen}, \bibinfo{author}{Z.~Zhu}, \bibinfo{author}{Z.~Hu} \& \bibinfo{author}{H.~Zeng},
 	\newblock \bibinfo{title}{{Semiconductor-topological insulator transition of two-dimensional SbAs induced by biaxial tensile strain}},
 	\newblock \emph{\bibinfo{journal}{Phys. Rev. B}} \textbf{\bibinfo{volume}{93}},
 	\bibinfo{pages}{245303} (\bibinfo{year}{2016}).
 	
 	 \bibitem{Winterfeld2013}
 	\bibinfo{author}{L.~Winterfeld}, \bibinfo{author}{L.~A.~Agapito}, \bibinfo{author}{J.~Li}, \bibinfo{author}{N.~Kioussis}, \bibinfo{author}{P.~Blaha} \& \bibinfo{author}{Y.~P.~Chen},
 	\newblock \bibinfo{title}{{Strain-induced topological insulator phase transition in HgSe}},
 	\newblock \emph{\bibinfo{journal}{Phys. Rev. B}} \textbf{\bibinfo{volume}{87}},
 	\bibinfo{pages}{075143} (\bibinfo{year}{2013}).
 	
 	\bibitem{Zhang2011}
 	\bibinfo{author}{X.~M.~Zhang}, \bibinfo{author}{W.~H.~Wang}, \bibinfo{author}{E.~K.~Liu}, \bibinfo{author}{G.~D.~Liu}, \bibinfo{author}{Z.~Y.~Liu} \& \bibinfo{author}{G.~H.~Wu},
 	\newblock \bibinfo{title}{{Influence of tetragonal distortion on the topological electronic structure of the half-Heusler compound LaPtBi from first principles}},
 	\newblock \emph{\bibinfo{journal}{Appl. Phys. Lett.}} \textbf{\bibinfo{volume}{99}},
 	\bibinfo{pages}{071901} (\bibinfo{year}{2011}).
 	
 	\bibitem{Teshome2019}
 	\bibinfo{author}{T.~Teshome} \& \bibinfo{author}{A.~Datta},
 	\newblock \bibinfo{title}{{Topological Phase Transition in $\mathrm{Sb_2Mg_3}$ Assisted by Strain}},
 	\newblock \emph{\bibinfo{journal}{ACS Omega}} \textbf{\bibinfo{volume}{4}},
 	\bibinfo{pages}{8701} (\bibinfo{year}{2019}).
 	
 	 \bibitem{Shao2017}
 	\bibinfo{author}{D.~Shao}, \bibinfo{author}{J.~Ruan}, \bibinfo{author}{J.~Wu}, \bibinfo{author}{T.~Chen}, \bibinfo{author}{Z.~Guo}, \bibinfo{author}{H.~Zhang}, \bibinfo{author}{J.~Sun}, \bibinfo{author}{L.~Sheng} \& \bibinfo{author}{D.~Xing},
 	\newblock \bibinfo{title}{{Strain-induced quantum topological phase transitions in $\mathrm{Na_3Bi}$}},
 	\newblock \emph{\bibinfo{journal}{Phys. Rev. B}} \textbf{\bibinfo{volume}{96}},
 	\bibinfo{pages}{075112} (\bibinfo{year}{2017}).
 	
 	 \bibitem{Tajkov2019}
 	\bibinfo{author}{Z.~Tajkov}, \bibinfo{author}{D.~Visontai}, \bibinfo{author}{L.~Oroszl\'{a}ny} \& \bibinfo{author}{J.~Koltai},
 	\newblock \bibinfo{title}{{Uniaxial strain induced topological phase transition in bismuth-–tellurohalide-–graphene heterostructures}},
 	\newblock \emph{\bibinfo{journal}{Nanoscale}} \textbf{\bibinfo{volume}{11}},
 	\bibinfo{pages}{12704} (\bibinfo{year}{2019}).
 	
 	\bibitem{Mutch2019}
 	\bibinfo{author}{J.~Mutch}, \bibinfo{author}{W.-C.~Chen}, \bibinfo{author}{P.~Went}, \bibinfo{author}{T.~Qian}, \bibinfo{author}{I.~Z.~Wilson}, \bibinfo{author}{A.~Adreev}, \bibinfo{author}{C.-C.~Chen} \& \bibinfo{author}{J.-H.~Chu},
 	\newblock \bibinfo{title}{{Evidence for a strain-tuned topological phase transition in $\mathrm{ZrTe_5}$}},
 	\newblock \emph{\bibinfo{journal}{Sci. Adv.}} \textbf{\bibinfo{volume}{5}},
 	\bibinfo{pages}{eaav9771} (\bibinfo{year}{2019}).
 	
 	\bibitem{Wang2017}
 	\bibinfo{author}{D.~Wang}, \bibinfo{author}{L.~Chen}, \bibinfo{author}{H.~Liu}, \bibinfo{author}{C.~Shi}, \bibinfo{author}{X.~Wang}, \bibinfo{author}{G.~Cui}, \bibinfo{author}{P.~Zhang} \& \bibinfo{author}{Y.~Chen},
 	\newblock \bibinfo{title}{{Strain induced band inversion and topological phase transition in methyl-decorated stanene film}},
 	\newblock \emph{\bibinfo{journal}{Sci. Rep.}} \textbf{\bibinfo{volume}{7}},
 	\bibinfo{pages}{17089} (\bibinfo{year}{2017}).
 	
 	\bibitem{LiuY2014}
 	\bibinfo{author}{Y.~Liu}, \bibinfo{author}{Y.~Y.~Li}, \bibinfo{author}{S.~Rajput}, \bibinfo{author}{D.~Gilks}, \bibinfo{author}{L.~Lari}, \bibinfo{author}{P.~L.~Gallindo}, \bibinfo{author}{M.~Weinert}, \bibinfo{author}{V.~K.~Lazarov} \& \bibinfo{author}{L.~Li},
 	\newblock \bibinfo{title}{{Tuning Dirac states by strain in the topological insulator $\mathrm{Bi_2Se_3}$}},
 	\newblock \emph{\bibinfo{journal}{Nat. Phys.}} \textbf{\bibinfo{volume}{10}},
 	\bibinfo{pages}{294} (\bibinfo{year}{2014}).
 	
 	\bibitem{Schindler2017}
 	\bibinfo{author}{C.~Schindler}, \bibinfo{author}{C.~Wiegand}, \bibinfo{author}{J.~Sichau}, \bibinfo{author}{L.~Tiemann}, \bibinfo{author}{K.~Nielsch}, \bibinfo{author}{R.~Zierold} \& \bibinfo{author}{R.~Blick},
 	\newblock \bibinfo{title}{{Strain-induced Dirac state shift in topological insulator $\mathrm{Bi_2Se_3}$ nanowires}},
 	\newblock \emph{\bibinfo{journal}{Appl. Phys. Lett.}} \textbf{\bibinfo{volume}{111}},
 	\bibinfo{pages}{171601} (\bibinfo{year}{2017}).
 	
 	\bibitem{Hwang2014}
 	\bibinfo{author}{J.~H.~Hwang}, \bibinfo{author}{S.~Kwon}, \bibinfo{author}{J.~Park}, \bibinfo{author}{J.~H.~Kim}, \bibinfo{author}{J.~Lee}, \bibinfo{author}{J.~S.~Kim}, \bibinfo{author}{H.-K.~Lyeo} \& \bibinfo{author}{J.~Y.~Park},
 	\newblock \bibinfo{title}{{Strain effects on in-plane conductance of the topological insulator $\mathrm{Bi_2Te_3}$}},
 	\newblock \emph{\bibinfo{journal}{Appl. Phys. Lett.}} \textbf{\bibinfo{volume}{104}},
 	\bibinfo{pages}{161613} (\bibinfo{year}{2014}).
 
 	\bibitem{Klotz2016}
 	\bibinfo{author}{J.~Klotz}, \bibinfo{author}{S.-C.~Wu}, \bibinfo{author}{C.~Shekhar}, \bibinfo{author}{Y.~Sun}, \bibinfo{author}{M.~Schmidt}, \bibinfo{author}{M.~Nicklas}, \bibinfo{author}{M.~Baenitz}, \bibinfo{author}{M.~Uhlarz}, \bibinfo{author}{J.~Wosnitza}, \bibinfo{author}{C.~Felser} \& \bibinfo{author}{B.~Yan},
 	\newblock \bibinfo{title}{{Quantum oscillations and the Fermi surface topology of the Weyl semimetal NbP}},
 	\newblock \emph{\bibinfo{journal}{Phys. Rev. B}} \textbf{\bibinfo{volume}{93}},
 	\bibinfo{pages}{121105(R)} (\bibinfo{year}{2016}).
 	
 	\bibitem{Lee2015}
 	\bibinfo{author}{C.-C.~Lee}, \bibinfo{author}{S.-Y.~Xu}, \bibinfo{author}{S.-M.~Huang}, \bibinfo{author}{D.~S.~Sanchez}, \bibinfo{author}{I.~Belopolski}, \bibinfo{author}{G.~Chang}, \bibinfo{author}{G.~Bian}, \bibinfo{author}{N.~Alidoust}, \bibinfo{author}{H.~Zheng}, \bibinfo{author}{M.~Neupane} \emph{et~al.},
 	\newblock \bibinfo{title}{{Fermi surface interconnectivity and topology in Weyl fermion semimetals TaAs, TaP, NbAs, and NbP}},
 	\newblock \emph{\bibinfo{journal}{Phys. Rev. B}} \textbf{\bibinfo{volume}{92}},
 	\bibinfo{pages}{235104} (\bibinfo{year}{2015}).
 	
 	\bibitem{Shekhar2015}
 	\bibinfo{author}{C.~Shekhar}, \bibinfo{author}{A.~N.~Nayak}, \bibinfo{author}{Y.~Sun}, \bibinfo{author}{M.~Schmidt}, \bibinfo{author}{M.~Nicklas}, \bibinfo{author}{I.~Leermakers}, \bibinfo{author}{U.~Zeitler}, \bibinfo{author}{Y.~Skourski}, \bibinfo{author}{J.~Wosnitza}, \bibinfo{author}{Z.~Liu} \emph{et~al.},
 	\newblock \bibinfo{title}{{Extremely large magnetoresistance and ultrahigh mobility in the topological Weyl semimetal candidate NbP}},
 	\newblock \emph{\bibinfo{journal}{Nat. Phys.}} \textbf{\bibinfo{volume}{11}},
 	\bibinfo{pages}{645} (\bibinfo{year}{2015}).
 	
 	\bibitem{Wang2016}
 	\bibinfo{author}{Z.~Wang}, \bibinfo{author}{Y.~Zheng}, \bibinfo{author}{Z.~Shen}, \bibinfo{author}{Y.~Lu}, \bibinfo{author}{H.~Fang}, \bibinfo{author}{F.~Sheng}, \bibinfo{author}{Y.~Zhou}, \bibinfo{author}{X.~Yang}, \bibinfo{author}{Y.~Li}, \bibinfo{author}{C.~Feng} \& \bibinfo{author}{Z.-A.~Xu},
 	\newblock \bibinfo{title}{{Helicity-protected ultrahigh mobility Weyl fermions in NbP}},
 	\newblock \emph{\bibinfo{journal}{Phys. Rev. B}} \textbf{\bibinfo{volume}{93}},
 	\bibinfo{pages}{121112(R)} (\bibinfo{year}{2016}).
 	
 	\bibitem{dosReis2016}
 	\bibinfo{author}{R.~D.~dos Reis}, \bibinfo{author}{S.-C.~Wu}, \bibinfo{author}{Y.~Sun}, \bibinfo{author}{M.~O.~Ajeesh}, \bibinfo{author}{C.~Shekhar}, \bibinfo{author}{M.~Schmidt}, \bibinfo{author}{C.~Felser}, \bibinfo{author}{B.~Yan} \& \bibinfo{author}{M.~Nicklas},
 	\newblock \bibinfo{title}{{Pressure tuning the Fermi surface topology of the Weyl semimetal NbP}},
 	\newblock \emph{\bibinfo{journal}{Phys. Rev. B}} \textbf{\bibinfo{volume}{93}},
 	\bibinfo{pages}{205102} (\bibinfo{year}{2016}).
 	
 	\bibitem{Wu2017}
 	\bibinfo{author}{S.-C.~Wu}, \bibinfo{author}{Y.~Sun}, \bibinfo{author}{C.~Felser} \& \bibinfo{author}{B.~Yan},
 	\newblock \bibinfo{title}{{Hidden type-II Weyl points in the Weyl semimetal NbP}},
 	\newblock \emph{\bibinfo{journal}{Phys. Rev. B}} \textbf{\bibinfo{volume}{96}},
 	\bibinfo{pages}{165113} (\bibinfo{year}{2017}).
 	
 	\bibitem{Liu2016}
 	\bibinfo{author}{Z.~K.~Liu}, \bibinfo{author}{L.~X.~Yang}, \bibinfo{author}{T.~Zhang}, \bibinfo{author}{H.~Peng}, \bibinfo{author}{H.~F.~Yang}, \bibinfo{author}{C.~Chen}, \bibinfo{author}{Y.~Zhang}, \bibinfo{author}{Y.~F.~Guo}, \bibinfo{author}{D.~Prabhakaran}, \bibinfo{author}{M.~Schmidt} \emph{et~al.},
 	\newblock \bibinfo{title}{{Evolution of the Fermi surface of Weyl semimetals in the transition metal pnictide family}},
 	\newblock \emph{\bibinfo{journal}{Nat. Mat.}} \textbf{\bibinfo{volume}{15}},
 	\bibinfo{pages}{27} (\bibinfo{year}{2015}).
 	
 	\bibitem{Belopolski2016}
 	\bibinfo{author}{I.~Belopolski}, \bibinfo{author}{S.-Y.~Xu}, \bibinfo{author}{D.~S.~Sanchez}, \bibinfo{author}{G.~Chang}, \bibinfo{author}{C.~Guo}, \bibinfo{author}{M.~Neupane}, \bibinfo{author}{H.~Zheng}, \bibinfo{author}{C.-C.~Lee}, \bibinfo{author}{S.-M.~Huang}, \bibinfo{author}{G.~Bian} \emph{et~al.},
 	\newblock \bibinfo{title}{{Criteria for Directly Detecting Topological Fermi Arcs in Weyl Semimetals}},
 	\newblock \emph{\bibinfo{journal}{Phys. Rev. Lett.}} \textbf{\bibinfo{volume}{116}},
 	\bibinfo{pages}{066802} (\bibinfo{year}{2016}).
 	
 	\bibitem{Souma2016}
 	\bibinfo{author}{S.~Souma}, \bibinfo{author}{Z.~Wang}, \bibinfo{author}{H.~Kotaka}, \bibinfo{author}{T.~Sato}, \bibinfo{author}{K.~Nakayama}, \bibinfo{author}{Y.~Tanaka}, \bibinfo{author}{H.~Kimizuka}, \bibinfo{author}{T.~Takahashi}, \bibinfo{author}{K.~Yamauchi}, \bibinfo{author}{T.~Oguchi} \emph{et~al.},
 	\newblock \bibinfo{title}{{Direct observation of nonequivalent Fermi-arc states of opposite surfaces in the noncentrosymmetric Weyl semimetal NbP}},
 	\newblock \emph{\bibinfo{journal}{Phys. Rev. B}} \textbf{\bibinfo{volume}{93}},
 	\bibinfo{pages}{161112(R)} (\bibinfo{year}{2016}).
 	
 	\bibitem{Zheng2016}
 	\bibinfo{author}{H.~Zheng}, \bibinfo{author}{S.-Y.~Xu}, \bibinfo{author}{G.~Bian}, \bibinfo{author}{C.~Guo}, \bibinfo{author}{G.~Chang}, \bibinfo{author}{D.~S.~Sanchez}, \bibinfo{author}{I.~Belopolski}, \bibinfo{author}{C.-C.~Lee}, \bibinfo{author}{S.-M.~Huang}, \bibinfo{author}{X.~Zhang} \emph{et~al.},
 	\newblock \bibinfo{title}{{Atomic-Scale Visualization of Quantum Interference on a Weyl Semimetal Surface by Scanning Tunneling Microscopy}},
 	\newblock \emph{\bibinfo{journal}{ACS Nano}} \textbf{\bibinfo{volume}{10}},
 	\bibinfo{pages}{1378} (\bibinfo{year}{2016}).
 	
 	\bibitem{Niemann2017}
 	\bibinfo{author}{A.~C.~Niemann}, \bibinfo{author}{J.~Gooth}, \bibinfo{author}{S.-C.~Wu}, \bibinfo{author}{S.~B\"{a}{\ss}ler}, \bibinfo{author}{P.~Sergelius}, \bibinfo{author}{R.~H\"{u}hne}, \bibinfo{author}{B.~Rellinghaus}, \bibinfo{author}{C.~Shekhar}, \bibinfo{author}{V.~S\"{u}{\ss}}, \bibinfo{author}{M.~Schmidt} \emph{et~al.},
 	\newblock \bibinfo{title}{{Chiral magnetoresistance in the Weyl semimetal NbP}},
 	\newblock \emph{\bibinfo{journal}{Sci. Rep.}} \textbf{\bibinfo{volume}{7}},
 	\bibinfo{pages}{43394} (\bibinfo{year}{2017}).
 	
 	\bibitem{Sergelius2016}
 	\bibinfo{author}{P.~Sergelius}, \bibinfo{author}{J.~Gooth}, \bibinfo{author}{S.~B\"{a}{\ss}ler}, \bibinfo{author}{R.~Zierold}, \bibinfo{author}{C.~Wiegand}, \bibinfo{author}{A.~C.~Niemann}, \bibinfo{author}{H.~Reith}, \bibinfo{author}{C.~Shekhar}, \bibinfo{author}{C.~Felser}, \bibinfo{author}{B.~Yan} \& \bibinfo{author}{K.~Nielsch},
 	\newblock \bibinfo{title}{{Berry phase and band structure analysis of the Weyl semimetal NbP}},
 	\newblock \emph{\bibinfo{journal}{Sci. Rep.}} \textbf{\bibinfo{volume}{6}},
 	\bibinfo{pages}{33859} (\bibinfo{year}{2016}).
 	
     \bibitem{Gooth2017}
	\bibinfo{author}{J.~Gooth}, \bibinfo{author}{A.~C.~Niemann}, \bibinfo{author}{T.~Meng}, \bibinfo{author}{A.~G.~Grushin}, \bibinfo{author}{K.~Landsteiner}, \bibinfo{author}{B.~Gotsmann}, \bibinfo{author}{F.~Menges}, \bibinfo{author}{M.~Schmidt}, \bibinfo{author}{C.~Shekhar}, \bibinfo{author}{V.~S\"{u}{\ss}} \emph{et~al.},
	\newblock \bibinfo{title}{{Experimental signatures of the mixed axial--gravitational anomaly in the Weyl semimetal NbP}},
	\newblock \emph{\bibinfo{journal}{Nature}} \textbf{\bibinfo{volume}{547}},
	\bibinfo{pages}{324} (\bibinfo{year}{2017}).
 	
 	\bibitem{Modic2019}
 	\bibinfo{author}{K.~A.~Modic}, \bibinfo{author}{T.~Meng}, \bibinfo{author}{F.~Ronning}, \bibinfo{author}{E.~D.~Bauer}, \bibinfo{author}{P.~J.~W.~Moll} \& \bibinfo{author}{B.~J.~Ramshaw},
 	\newblock \bibinfo{title}{{Thermodynamic Signatures of Weyl Fermions in NbP}},
 	\newblock \emph{\bibinfo{journal}{Sci. Rep.}} \textbf{\bibinfo{volume}{9}},
 	\bibinfo{pages}{2095} (\bibinfo{year}{2019}).
 	
 	\bibitem{Neubauer2018}
 	\bibinfo{author}{D.~Neubauer}, \bibinfo{author}{A.~Yaresko}, \bibinfo{author}{W.~Li}, \bibinfo{author}{A.~L\"{o}hle}, \bibinfo{author}{R.~H\"{u}bner}, \bibinfo{author}{M.~B.~Schilling}, \bibinfo{author}{C.~Shekhar}, \bibinfo{author}{C.~Felser}, \bibinfo{author}{M.~Dressel} \& \bibinfo{author}{A.~V.~Pronin},
 	\newblock \bibinfo{title}{{Optical conductivity of the Weyl semimetal NbP}},
 	\newblock \emph{\bibinfo{journal}{Phys. Rev. B}} \textbf{\bibinfo{volume}{98}},
 	\bibinfo{pages}{195203} (\bibinfo{year}{2018}).
 	
 	\bibitem{Xu2017}
 	\bibinfo{author}{J.~Xu}, \bibinfo{author}{D.~E.~Bugaris}, \bibinfo{author}{Z.~L.~Xiao}, \bibinfo{author}{Y.~L.~Wang}, \bibinfo{author}{D.~Y.~Chung}, \bibinfo{author}{M.~G.~Kanatzidis} \& \bibinfo{author}{W.~K.~Kwok},
 	\newblock \bibinfo{title}{{Reentrant metallic behavior in the Weyl semimetal NbP}},
 	\newblock \emph{\bibinfo{journal}{Phys. Rev. B}} \textbf{\bibinfo{volume}{96}},
 	\bibinfo{pages}{115152} (\bibinfo{year}{2017}).
 	
 	\bibitem{Zheng2017}
 	\bibinfo{author}{H.~Zheng}, \bibinfo{author}{G.~Chang}, \bibinfo{author}{S.-M.~Huang}, \bibinfo{author}{C.~Guo}, \bibinfo{author}{X.~Zhang}, \bibinfo{author}{S.~Zhang}, \bibinfo{author}{J.~Yin}, \bibinfo{author}{S.-Y.~Xu}, \bibinfo{author}{I.~Belopolski}, \bibinfo{author}{N.~Alidoust} \emph{et~al.},
 	\newblock \bibinfo{title}{{Mirror Protected Dirac Fermions on a Weyl Semimetal NbP Surface}},
 	\newblock \emph{\bibinfo{journal}{Phys. Rev. Lett.}} \textbf{\bibinfo{volume}{119}},
 	\bibinfo{pages}{196403} (\bibinfo{year}{2017}).
 	
 	\bibitem{Chang2016}
 	\bibinfo{author}{G.~Chang}, \bibinfo{author}{S.-Y.~Xu}, \bibinfo{author}{H.~Zheng}, \bibinfo{author}{C.-C.~Lee}, \bibinfo{author}{S.-M.~Huang}, \bibinfo{author}{I.~Belopolski}, \bibinfo{author}{D.~S.~Sanchez}, \bibinfo{author}{G.~Bian}, \bibinfo{author}{N.~Alidoust}, \bibinfo{author}{T.-R.~Chang} \emph{et~al.},
 	\newblock \bibinfo{title}{{Signatures of Fermi Arcs in the Quasiparticle Interferences of the Weyl Semimetals TaAs and NbP}},
 	\newblock \emph{\bibinfo{journal}{Phys. Rev. Lett.}} \textbf{\bibinfo{volume}{116}},
 	\bibinfo{pages}{066601} (\bibinfo{year}{2016}).
 	
 	\bibitem{Sun2015}
 	\bibinfo{author}{Y.~Sun}, \bibinfo{author}{S.-C.~Wu} \& \bibinfo{author}{B.~Yan}
 	\newblock \bibinfo{title}{{Topological surface states and Fermi arcs of the noncentrosymmetric Weyl semimetals TaAs, TaP, NbAs, and NbP}}.
 	\newblock \emph{\bibinfo{journal}{Phys. Rev. B}} \textbf{\bibinfo{volume}{92}},
 	\bibinfo{pages}{115428} (\bibinfo{year}{2015}).
 	
 	\bibitem{Weng2015}
 	\bibinfo{author}{H.~Weng}, \bibinfo{author}{C.~Fang}, \bibinfo{author}{Z.~Fang}, \bibinfo{author}{B.~A.~Bernevig} \& \bibinfo{author}{X.~Dai},
 	\newblock \bibinfo{title}{{Weyl Semimetal Phase in Noncentrosymmetric Transition-Metal Monophosphides}},
 	\newblock \emph{\bibinfo{journal}{Phys. Rev. X}} \textbf{\bibinfo{volume}{5}},
 	\bibinfo{pages}{011029} (\bibinfo{year}{2015}).
 	
 	\bibitem{Huang2015}
 	\bibinfo{author}{S.-M.~Huang}, \bibinfo{author}{S.-Y.~Xu}, \bibinfo{author}{I.~Belopolski}, \bibinfo{author}{C.-C.~Lee}, \bibinfo{author}{G.~Chang}, \bibinfo{author}{B.~Wang}, \bibinfo{author}{N.~Alidoust}, \bibinfo{author}{G.~Bian}, \bibinfo{author}{M.~Neupane}, \bibinfo{author}{C.~Zhang} \emph{et~al.},
 	\newblock \bibinfo{title}{{A Weyl Fermion semimetal with surface Fermi arcs in the transition metal monopnictide TaAs class}},
 	\newblock \emph{\bibinfo{journal}{Nat. Comm.}} \textbf{\bibinfo{volume}{6}},
 	\bibinfo{pages}{7373} (\bibinfo{year}{2015}).
 	
 	\bibitem{Greenblatt1996}
 	\bibinfo{author}{J.~Xu}, \bibinfo{author}{J.~M.~Greenblatt}, \bibinfo{author}{T.~Emge} \& \bibinfo{author}{P.~H\"{o}hn},
 	\newblock \bibinfo{title}{{Crystal Structure, Electrical Transport, and Magnetic Properties of Niobium Monophosphide}},
 	\newblock \emph{\bibinfo{journal}{Inorg. Chem.}} \textbf{\bibinfo{volume}{35}},
 	\bibinfo{pages}{845} (\bibinfo{year}{1996}).
 	
 	\bibitem{Hicks2014}
 	\bibinfo{author}{C.~W.~Hicks}, \bibinfo{author}{M.~E.~Barber}, \bibinfo{author}{S.~D.~Edkins}, \bibinfo{author}{D.~O.~Brodsky} \& \bibinfo{author}{A.~P.~Mackenzie},
 	\newblock \bibinfo{title}{{Piezoelectric-based apparatus for strain tuning}},
 	\newblock \emph{\bibinfo{journal}{Rev. Sci. Instrum.}} \textbf{\bibinfo{volume}{85}},
 	\bibinfo{pages}{065003} (\bibinfo{year}{2014}).
 	
 		
 	\bibitem{Steppke2017}
 	\bibinfo{author}{A.~Steppke}, \bibinfo{author}{L.~Zhao}, \bibinfo{author}{M.~E.~Barber}, \bibinfo{author}{T.~Scaffidi}, \bibinfo{author}{F.~Jerzembeck}, \bibinfo{author}{H.~Rosner}, \bibinfo{author}{A.~S.~Gibbs}, \bibinfo{author}{Y.~Maeno}, \bibinfo{author}{S.~H.~Simon}, \bibinfo{author}{A.~P.~Mackenzie} \& \bibinfo{author}{C.~W.~Hicks},
 	\newblock \bibinfo{title}{{Strong peak in $T_\mathrm{C}$ of $\mathrm{Sr_2RuO_4}$ under uniaxial pressure}},
 	\newblock \emph{\bibinfo{journal}{Science}} \textbf{\bibinfo{volume}{13}},
 	\bibinfo{pages}{eaaf9398} (\bibinfo{year}{2017}).
 	
 	\bibitem{Jo2019}
 	\bibinfo{author}{N.~H.~Jo}, \bibinfo{author}{L.-L.~Wang}, \bibinfo{author}{P.~P.~Orth}, \bibinfo{author}{S.~Bud'ko} \& \bibinfo{author}{P.~C.~Canfield},
 	\newblock \bibinfo{title}{{Magnetoelastoresistance in $\mathrm{WTe_2}$: Exploring electronic structure and extremely large magnetoresistance under strain}},
 	\newblock \emph{\bibinfo{journal}{Proc. Natl. Acad. Sci. USA}} \textbf{\bibinfo{volume}{116}},
 	\bibinfo{pages}{25524} (\bibinfo{year}{2019}).
 	
    \bibitem{kresse1996}
    \bibinfo{author}{G.~Kresse} \& \bibinfo{author}{J.~Furthm{\"{u}}ller},
    \newblock \bibinfo{title}{{Efficiency of ab-initio total energy calculations
    for metals and semiconductors using a plane-wave basis set}},
    \newblock \emph{\bibinfo{journal}{Computational Materials Science}}
    \textbf{\bibinfo{volume}{6}}, \bibinfo{pages}{15} (\bibinfo{year}{1996}).
    
    \bibitem{perdew1996}
 	\bibinfo{author}{J.~P.~Perdew}, \bibinfo{author}{K.~Burke} \& \bibinfo{author}{M.~Ernzerhof},
 	\newblock \bibinfo{title}{{Generalized Gradient Approximation Made Simple}},
 	\newblock \emph{\bibinfo{journal}{Phys. Rev. Lett.}} \textbf{\bibinfo{volume}{77}},
 	\bibinfo{pages}{3865} (\bibinfo{year}{1996}).
    
    \bibitem{Mostofi2008}
	\bibinfo{author}{A.~A.~Mostofi}, \bibinfo{author}{J.~R.~Yates}, \bibinfo{author}{Y.-S.~Lee}, \bibinfo{author}{I.~Souza}, \bibinfo{author}{D.~Vanderbilt} \& \bibinfo{author}{N.~Marzari},
	\newblock \bibinfo{title}{{wannier90: A tool for obtaining maximally-localised Wannier functions}},
	\newblock \emph{\bibinfo{journal}{Comp. Phys. Comm.}} \textbf{\bibinfo{volume}{178}},
	\bibinfo{pages}{685} (\bibinfo{year}{2008}).
     	
    \bibitem{Liu2017}
    \bibinfo{author}{L.~Liu}, \bibinfo{author}{Z.-Q.~Wang}, \bibinfo{author}{C.-E.~Hu}, \bibinfo{author}{Y.~Cheng} \& \bibinfo{author}{G.-F.~Ji},
    \newblock \bibinfo{title}{{Comparative study on structural, elastic, dynamical, and thermodynamic properties of Weyl semimetals MX (M = Ta or Nb; X = As or P)}},
    \newblock \emph{\bibinfo{journal}{Solid State Commun.}} \textbf{\bibinfo{volume}{263}},
    \bibinfo{pages}{10} (\bibinfo{year}{2017}).
    
 	\bibitem{Shoenberg}
 	\bibinfo{author}{D.~Shoenberg},
 	\newblock \emph{\bibinfo{title}{{Magnetic Oscillations in Metals}}}
 	(\bibinfo{publisher}{Cambridge Univ. Press}, \bibinfo{year}{1984}).
	
	\bibitem{Audourd2018}
	\bibinfo{author}{A.~Audouard} \& \bibinfo{author}{J.~Fortin},
	\newblock \bibinfo{title}{{Does Fourier analysis yield reliable amplitudes of quantum oscillations?}},
	\newblock \emph{\bibinfo{journal}{EPJ Appl. Phys.}} \textbf{\bibinfo{volume}{83}},
	\bibinfo{pages}{30201} (\bibinfo{year}{2018}).
	
	 \bibitem{Mercure2008}
	\bibinfo{author}{J.-F.~Mercure},
	\newblock \emph{\bibinfo{title}{{The de Haas van Alphen effect near a quantum critical end point in $\mathrm{Sr_3Ru_2O_7}$}}}
	(\bibinfo{address}{Univ. of St. Andrews}, \bibinfo{year}{2008}), \bibinfo{url}{https://arxiv.org/abs/1001.3980}.

	\bibitem{Bedoya2020}
	\bibinfo{author}{A.~Bedoya-Pinto}, \bibinfo{author}{A.~K.~Pandeya}, \bibinfo{author}{D.~Liu}, \bibinfo{author}{H.~Deniz},
	\bibinfo{author}{K.~Chang},
	\bibinfo{author}{H.~Tan},
	\bibinfo{author}{H.~Han},
	\bibinfo{author}{J.~Jena},
	\bibinfo{author}{I.~Kostanovskiy}
	\& \bibinfo{author}{S.~P.~Parkin},
	\newblock \bibinfo{title}{{Realization of Epitaxial NbP and TaP Weyl Semimetal Thin Films}},
	\newblock \emph{\bibinfo{journal}{ACS Nano}} \textbf{\bibinfo{volume}{14}},
	\bibinfo{pages}{4405} (\bibinfo{year}{2020}).
	

\end{thebibliography}
\end{document}